\documentstyle[aps,pre,citesort,twocolumn,psfig,amsmath,amssymb]{revtex}

\begin{document} 

\twocolumn[\hsize\textwidth\columnwidth\hsize\csname 
@twocolumnfalse\endcsname


\title{Onset of Convection in a Very Compressible Fluid : The Transient Toward
Steady State}

\author{Horst Meyer and Andrei B. Kogan*}

\address{Department of Physics, Duke University, Durham, NC 27708-0305.
\\}

\address{*present address: Department of Physics, MIT, Cambridge, MA 
02139-4307.\\}


\maketitle

\begin{abstract}

We analyze the time  profile  $\Delta T(t)$ of the temperature difference,
measured across a very compressible supercritical $^3$He fluid layer in its
convective state. The experiments were done along the critical isochore in a
Rayleigh-B\'{e}nard cell after starting the vertical constant heat 
flow $q$. For $q$ sufficiently well above that needed  for the convection onset, the transient $\Delta T(t)$ for a given $\epsilon\equiv(T-T_c)/T_c$, with $T_c$ = 3.318K, shows a damped oscillatory profile with period $t_{osc}$ modulating a smooth base profile. The smooth profile forms the exponential tail of the transient which tends to  the steady-state $\Delta T(\infty)$ with a
time constant $\tau_{tail}$. The scaled times $t_{osc}/t_D$ and $\tau_{tail}/t_D$ from all the data could be collapsed onto two curves as a function of the Rayleigh number over $\sim$ 3.5 decades. Here $t_D$ is the characteristic thermal diffusion time. Furthermore comparisons are made between  measurements of a third characteristic time $t_m$ between the first peak and the first minimum in the $\Delta T(t)$  profile and its  estimation by Onuki et al. Also comparisons are made  between the observed oscillations and the 2D simulations by Onuki et al.  and by Amiroudine and Zappoli. For $\epsilon < 9\times 10^{-3}$ the experiments show a crossover to a different transient regime. This new regime, which we briefly  describe, is not understood at present.

\vspace{-0.6cm}\\

\end{abstract}

\narrowtext
\vskip2pc]

\section{Introduction}

Many common fluids can be treated as incompressible, i.e.  one can assume that
their  density, while perhaps varying slightly with temperature, is
pressure-independent. This is one of the assumptions of   the so 
called Boussinesq
approximation\cite{Tritton:88}. Throughout this paper, we refer to a fluid as
``compressible" if the non-zero value of the  isothermal expansion coefficient
$\beta_T = 1/\rho (\partial \rho/\partial P)_T$ significantly 
influences the fluid
dynamics and thus cannot be neglected. Here $\rho$ is the density and $P$ the
pressure.

The increased interest in understanding the properties of  a 
compressible flow is
driven by both experimental and theoretical work on highly developed turbulence
(See the review article by L.P. Kadanoff\cite{Kadanoff:2001} and references
therein).  Compressible supercritical fluids allow to explore flow 
regimes which
are not easily accessible experimentally: for example, Prandtl 
numbers $Pr$ of the
order of 500 can be obtained, while for an ordinary  gas $Pr$ is of order
$1$\cite{Tritton:88}.

  A standard experimental approach to study flow is to use a 
Rayleigh-B\'{e}nard (RB)
geometry, i.e.  a laterally confined horizontal layer of fluid 
filling the space
between two horizontal surfaces 
\cite{Ashkenazi:S:99,Niemela:2000,Chavanne:2001}.
In a typical experiment, one  applies externally heat with flux 
density $q$, and
measures the temperature difference $\Delta T(t)$ across the layer. 
The observable
quantity of interest to theory is the steady state value $\Delta 
T(\infty)$ as a
function of $q$. If the fluid properties are known, the relations
$Nu$($Ra$,$Pr$) can be obtained and compared with predictions. Here 
$Nu$ and $Ra$
are the Nusselt and Rayleigh numbers, which represent the normalized 
heat current
and temperature difference, respectively.

Generally, a compressible fluid does not obey the Boussinesq approximation.
However, in a Rayleigh-B\'{e}nard geometry,  the Boussinesq 
approximation is always
valid provided that the layer height is sufficiently small given the value of
$\beta_T$. And vice versa, even a weakly compressible fluid will violate  the
Boussinesq approximation if the layer thickness is large 
enough\cite{Landau:L:}.
Therefore, while the effects due to finite $\beta_T$ are interesting 
on their own,
they can also pose a challenge in understanding results of 
experiments focusing on
other phenomena: for example, studies of  highly turbulent flow are 
often performed
in tall RB  cells, which is desirable because of the rapid  $(\sim 
h^3)$ increase
of the Rayleigh number with the cell height.

We are interested in  the effects due to non-vanishing $\beta_T$, and  we  use
supercritical $^3$He as a test fluid. Supercritical fluids provide
a unique opportunity to study the compressibility-related effects, because
$\beta_T$ exibits a rapid  divergence at the critical point. As a 
result, one can
continuously change the system from a weakly-compressible state to a 
regime where
the dynamics is dominated by the high compressibility. The analysis 
presented here
extends that presented in an earlier paper \cite{Kogan:M:2001}  and 
focuses  on the
time-dependent transient $\Delta T(t)$. As explained there in detail, 
the apparatus
design was  chosen so that  the high compressibility is the only significant
deviation from the set  of assumptions required by the Boussinesq 
approximation.

Even in the absence of convection, e.g. under zero gravity 
conditions, the dynamics
of a highly compressible fluid can be qualitatively different  from an
incompressible system. As was first predicted theoretically by Onuki 
and Ferrell
\cite{Onuki:F:90}, and then later demonstrated experimentally by several
groups\cite{Beysens}, the density-temperature dynamics following a small
disturbance of the temperature of the walls depends on whether the 
pressure $P$  or
the total volume  V ( in other words, the space-average density 
$<\rho>$ ) is kept
constant. In the latter case, the mechanism called ``Piston Effect" 
dominates the
dynamics: Following a small positive step in the wall temperature, 
sharp gradients
of
$\rho$ and
$T$ form quickly at the walls and produce ``boundary layers", which expand away
from the walls into the bulk of the fluid. The ``piston" time scale for the
formation of the boundary layer is typically seconds, while the 
relaxation back to
the uniform density distribution happen on the diffusive time scale which in
practice is minutes to hours. These boundary layers compress the bulk fluid
adiabatically, which gives  rise to an apparent fast relaxation of  the bulk
temperature:  it becomes correlated with the wall temperature on the 
piston time
scale instead of that of the diffusion. At constant $P$, or if the 
compressibility
is small,  the density field does not respond to the  wall 
temperature variations,
no boundary layers are formed and the temperature undergoes a conventional
diffusive relaxation according to the Fourier Law.

If the fluid is subjected to the earth gravity field $g$, the most basic
compressibility- related effect is the change in the conditions for 
the mechanical
stability.
  For a supercritical fluid,  the compressibility increases as
$(T-T_c)^{-\gamma_{eff}}$ with $\gamma_{eff} \approx$ 1.19 for $^3$He,   as the
critical point is approached from above along the critical isochore. 
As a result,
  the onset transition  to the convective state exhibits a crossover 
from the usual
Rayleigh- to the Schwarzschild criterion. Far above
$T_c$, the temperature drop across the fluid layer of height $L$ at 
the convection
onset transition is given by $\Delta T_{ons}$=$\Delta T_{Ra}$ while close to
$T_c$, the stability regime is determined  by the ``adiabatic 
temperature gradient"
$\nabla T|_{ad}$\cite{Tritton:88} so that
$\Delta T_{ad}$ = $\nabla T|_{ad}\times L$.  Here
$\Delta T_{Ra}$, the onset temperature for the Rayleigh criterion alone and
$\nabla T|_{ad}$ are given by
\begin{equation}
     \Delta T_{Ra} = Ra_c(D_T\eta/g\rho \alpha_P L^3)
\end{equation}
   and
\begin{equation}
\nabla T|_{ad}\equiv\left(\frac{\partial T}{\partial 
p}\right)_s\frac{dp}{dz}=\rho
g T
\alpha_P/C_P =\rho g(1-\frac{C_V}{C_P})(\frac{\partial T}{\partial p})_{\rho}.
\end{equation}
     Here $\rho$, g, $\eta$,  $\alpha_P$, $D_T$ and $C_P$  are respectively the
density, gravitational acceleration, shear viscosity, coefficient of isobaric
thermal expansion, thermal diffusivity and isobaric specific 
heat\cite{adiabatic}.
For cells with large aspect ratio,   the theoretical value of the 
critical Rayleigh
number
$Ra_c$ is $\approx$ 1708. In general, as has been shown previously,
\cite{Gitterman:S:71,Carles:U:99}
\begin{equation}
\Delta T_{ons} = \Delta T_{ad} + \Delta T_{Ra}
\label{eq:ons}
\end{equation}
   A ``potential temperature" $\theta$ can be defined as
\begin{equation}
     \frac{\partial\theta}{\partial z} = \frac{\partial T}{\partial z} -\nabla
T|_{ad}
\end{equation}
   where $z$ is the vertical coordinate (See for instance 
ref.\cite{Tritton:88}).
When
$\theta$ is substituted in place of
$T$ into the hydrodynamic equations for a compressible fluid, the 
Boussinesq form
is recovered, provided the large $\nabla T|_{ad}$ causes the only 
deviation from the
Boussinesq set of criteria (A,B,C and D in Appendix to Chapter 14 in
ref\cite{Tritton:88}). This substitution leads to the corrected forms
$Nu^{corr}$ and $Ra^{corr}$ where account is taken for the effect of
compressibility. Here
\begin{equation}
     Ra^{corr} = \frac{Ra(\Delta T - \Delta T_{ad})}{\Delta T}{\rm
,}\,\,Nu^{corr} =
\frac{(\Delta T_{diff} - \Delta T_{ad})}{(\Delta T -\Delta T_{ad})}
\end{equation}
   with $Ra \equiv \alpha_P \rho g L^3\Delta T/\eta D_T$. Also $\Delta 
T_{diff}$ is
the temperature drop across the fluid in the diffusive regime for the same heat
current
$q$ that produces the observed
$\Delta T$. For a discussion of these corrected numbers, we refer to
ref\cite{Kogan:M:2001}.

Very recently, we reported  a study of the convection in  supercritical $^3$He.
\cite{Kogan:M:2001}. The experiments were carried out in a 
Rayleigh-B\'{e}nard  cell
with a layer thickness $L$=0.106 cm and  aspect ratio of 57. The top 
boundary was
kept at a constant temperature. A constant heating power  was applied 
at the bottom
plate, the temperature of which was left floating. Both the top and the bottom
surfaces  had a high thermal conductivity and thus there was no 
lateral temperature
inhomogeneity. A temperature difference
$\Delta T(t)$ was measured across the fluid layer as a function of 
time $t$ after
the heat current was started. Its value in the steady state is 
henceforth denoted
as $\Delta T$ or as $\Delta T(\infty)$.

The cell was filled to the critical density $\rho =
\rho_c$ and then sealed, thus the experiments were always done under 
fixed volume
conditions. The reduced temperature $\epsilon = (T - T_c)/T_c$ , 
where $T_c$=3.318K
is the critical temperature,  was varied over the range
0.2$\geq\epsilon\geq 5\times 10^{-4}$, where the fluid properties
changed substantially, for instance $1.2\times
10^{-6}<\beta_T<1.5\times 10^{-3}$ cm$^2$/dyne  and  $2<Pr< 590 $.
The effects of gravity on the vertical density distribution were at most of the
order of a few \% at the temperatures closest to $T_c$. The heat currents
$q$ across the fluid were kept small enough that  there were no 
significant changes
of the fluid properties across the fluid layer, even those strongly 
dependent on
$\epsilon$ and on $\rho$.  The fluid system was therefore never 
highly non-linear
in the dependence of properties on $z$ in the cell. In addition, as mentioned
earlier, all the assumptions required for the validity of the Boussinesq
approximation, except the compressibility requirements, were fulfilled. For the
numerical estimates of the maximum deviations as well as  the experimental
procedures and details, see ref~\cite{Kogan:M:2001}.

Over the range 0.009$\leq\epsilon\leq$ 0.2, patterns with damped 
oscillations were
observed. Their frequency $t_{osc}^{-1}$ increased with $q$ and eventually the
oscillations were smoothed out (Fig.9 of ref.\cite{Kogan:M:2001}). For
$\epsilon \leq $0.009, the $\Delta T(t)$ profiles changed appreciably 
from those
just described. These unusual transients  stimulated theoretical studies of
this supercritical fluid system, undertaken independently by Amiroudine and
Zappoli (AZ)\cite{Amiroudine:2000,Amiroudine:Z:2002} and by Onuki and
coworkers\cite{Chiwata:O:2001,Furukawa:O:2002}. Both groups carried 
out simulation
studies with different codes and approximations, and some of their respective
results will be discussed in this paper in connection with our data 
analysis.  The
simulations used the Navier Stokes equation and the heat conduction 
equation where
a term was added in both, which takes into account the compressibility as
represented by   the adiabatic temperature gradient. The simulations 
were in 2D, and
in ref.\cite{Chiwata:O:2001} the Stokes approximation was used, where 
in the N-S
equation the time derivative of the velocity was set to zero. The fluid was
contained in a cell with an aspect ratio of 4 with vertical periodic boundary
conditions. In ref. \cite{Furukawa:O:2002}, the Stokes approximation 
was not used
and the  simulations then extended to $Ra^{corr}-Ra_c  = 4 \times 10^6$. In the
work by AZ\cite{Amiroudine:Z:2002},  the simulations were carried out up to
$Ra^{corr}-Ra_c  = 6 \times 10^6$ with the use of a supercomputer. Here various
aspect ratios were used, but always with periodic boundary 
conditions. Furthermore
the instability onset time of the same fluid, after the  heat current 
start, was
theoretically investigated as a function of
$q$ and $\epsilon$ by Carl\`{e}s\cite{Carles:2000} and by  El Khoury and
Carl\`{e}s\cite{Carles:2002}.

To describe a transient, we introduce  several characteristic  time 
scales that can
be extracted from recorded data.  Two of these, the period of damped 
oscillations
$t_{osc}$ and the relaxation time $\tau_{tail}$ of the transient tail 
end, give in
scaled form a particularly interesting universal result,  which has not been
predicted so far. In Section II, we give an overview of various types 
of  profiles
$\Delta T(t)$, and the conditions under which  they are observed, and 
we define the
characteristic times used in the analysis. This is followed in 
Section III by the
scaled presentation of
$t_{osc}$ and $\tau_{tail}$ versus the Rayleigh number. In Section IV, another
measured characteristic time in the $\Delta T(t)$ profile, $t_m$ as defined by
Chiwata and Onuki\cite{Chiwata:O:2001}, is compared with their predictions and
simulations. Also the simulation results by AZ are discussed.
   The temperature region for
$\epsilon < 9\times 10^{-3}$, where no oscillations were observed, is briefly
described. Finally  Section V presents the main results and conclusions.

\section{Convection Onset and Transient profiles}.

   We first give an overview of the various profiles $\Delta T(t)$ 
observed  after
the start of the constant heat flow
$q$  at $t$ = 0, and we  find that they exhibit several
distinct characteristic shapes. For each of these, one can identify an
approximate range in the parameters  $\epsilon$ and $\Delta T(\infty)$, or a
``zone"   where the shape is observed.  We first describe the typical 
$\Delta T(t)$
shapes and the corresponding ``zones" in the $\Delta 
T(\infty)-\epsilon$  plane.
Then, we define two characteristic times and discuss their respective
uncertainties. Because the
$\Delta T(t)$ observed above and below
$\epsilon \approx 9\times10^{-3}$ are quite different, we consider 
them separately.

\begin{figure}[htb]
\center{\parbox{3.375in}{\psfig{file=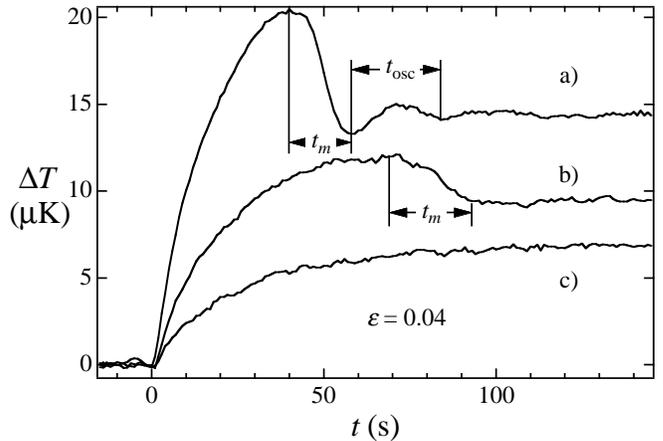,width=3.375in}}}
\caption{Recordings $\Delta T(t)$ after starting the heat current, with the
definition of the times $t_{osc}$ (Section III) and $t_m$ (Section IV). The
recordings were made at
$\epsilon$ = 0.04 with the heat flow $q$ (in $10^{-8}$ W/cm$^2$ a) 5.17 b)
2.53, c) 1.21.}
\end{figure}

\begin{figure}[htb]
\center{\parbox{3.375in}{\psfig{file=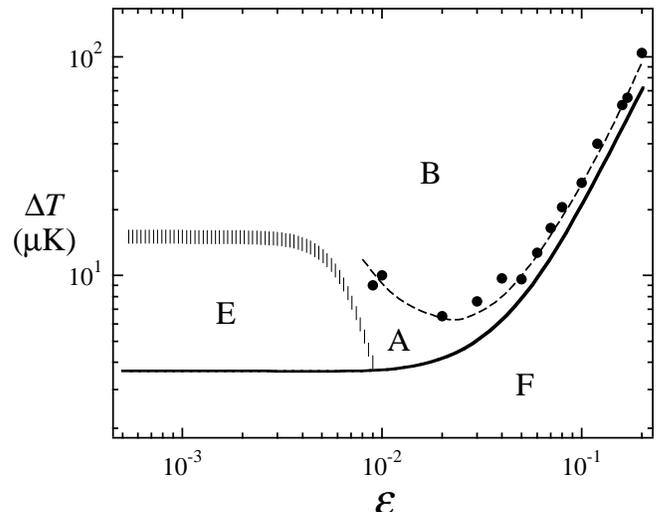,width=3.375in}}}
\caption{Tentative ``zone diagram" for the profiles $\Delta T(t)$ in 
the $\Delta
T-\epsilon$ plane. The line of solid circles forms the boundary 
between zone A and
B. There are no data for this boundary below $\epsilon < 
9\times10^{-3}$, where no
oscillations are observed. In zone E, no overshoot is observed.}
\end{figure}

\subsection {Region for $\epsilon \geq 9\times10^{-3}$}

In the non-convective regime, the transient $\Delta T(t)$  follows 
the prediction
of the Fourier Law, modified to accomodate the adiabatic - or ``Piston"
effect.  An example trace can be seen in Fig.1c.  $\Delta T(t)$ tends
exponentially to  $\Delta T(\infty)$  with a relaxation time determined by the
thermal diffusivity $D_T$
\cite{Behringer:O:M:90}. This corresponds to zone $F$ of the diagram 
in Fig.2. The
solid line  separating zone $F$ from the rest of the field is  given by the
condition of the mechanical stability (Eq. \ref{eq:ons}) confirmed 
experimentally
in \cite{Kogan:M:2001}.

When $q$ is increased above the value that corresponds to the 
convection onset, the
transient $\Delta T(t)$ exibits an initial rise followed by an 
overshoot. This is a
well-known effect which can be understood  as a result of the fluid inertia
\cite{Behringer:A:77,Behringer:85}: the convection doesn't start 
immediately after
the
  mechanical stability condition is violated. Instead,
  $\Delta T(t)$ rises in accordance with the  non-convective dynamics past the
convection onset threshold until the fluid flow develops.   Two 
examples of such
behavior are shown at very different values of
$\epsilon$ in  Fig.1b and Fig.3b.  This behavior  is observed in zone 
A of Fig.2.
Interestingly, we do not generally find  the relaxation towards  the $\Delta
T(\infty)$  to be exponential, as predicted and experimentally observed in the
immediate vicinity of the onset for Boussinesq fluids
\cite{Behringer:A:77,Behringer:85}.  Instead, we find an almost linear decrease
followed by a kink in the  transient. The noise in Fig.1 obscures the 
``kink" which
is clearly seen in Fig.3b at the end of the overshoot. We henceforth 
refer to the
transients of this type as a ``truncated oscillation".

A further increase in $q$  produces damped oscillations in $\Delta T(t)$ with a
period
$t_{osc}$ (Figs 1a and 3a). Most of the analysis presented in this 
work focuses on
this regime. These oscillations  are observed in the lower part of 
zone B in Fig.2
above its boundary with zone A.  For $\epsilon < 9 \times10^{-3}$, 
this boundary
is undetermined because of lack of data.

\begin{figure}[htb]
\center{\parbox{3.375in}{\psfig{file=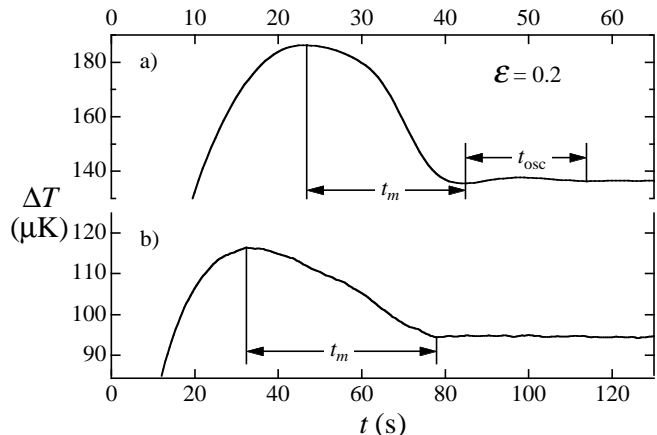,width=3.375in}}}
\caption{Recordings $\Delta T(t)$ at $\epsilon$ = 0.2, with the 
definition of the
times $t_m$ and $t_{osc}$. a) In zone B of fig.2 with damped 
oscillations and b) in
zone A with truncated oscillations. The heat flow $q$ (in
$10^{-7}$ W/cm$^2$) is a) 3.89, b) 2.16.}
\end{figure}

With increasing $q$ in zone B, the profile $\Delta T(t)$ becomes more 
complex but
has two  components amenable to analysis:  damped oscillations of period
$t_{osc}(q,\epsilon)$ followed by  a smooth base profile with a minimum, beyond
which
$\Delta T(t)$ tends exponentially to $\Delta T(\infty)$ from below 
with a relaxation
time $\tau_{tail}(q,\epsilon)$. This evolution with $q$ at
$\epsilon$ = 0.05 can be seen in Fig.4, where for emphasis only the 
top portion of
the
$\Delta T(t)$ profile is shown. The visibility and relative amplitude of these
components are functions of $\epsilon$ and of $q$. For a given $\epsilon$, both
$\tau_{tail}$ and $t_{osc}$ decrease with increasing $q$, and always $t_{osc} <
\tau_{tail}$. When $t_{osc}$ becomes comparable to the time constant of the
thermometer circuitry, $\simeq$ 1.3 s., the oscillation amplitude and 
eventually
the very first peak become smoothed out and only the base profile relaxation is
observed. Hence in a portion of the recorded traces, either $t_{osc}$ or
$\tau_{tail}$ can be determined, but not both.

\begin{figure}[htb]
\center{\parbox{3.375in}{\psfig{file=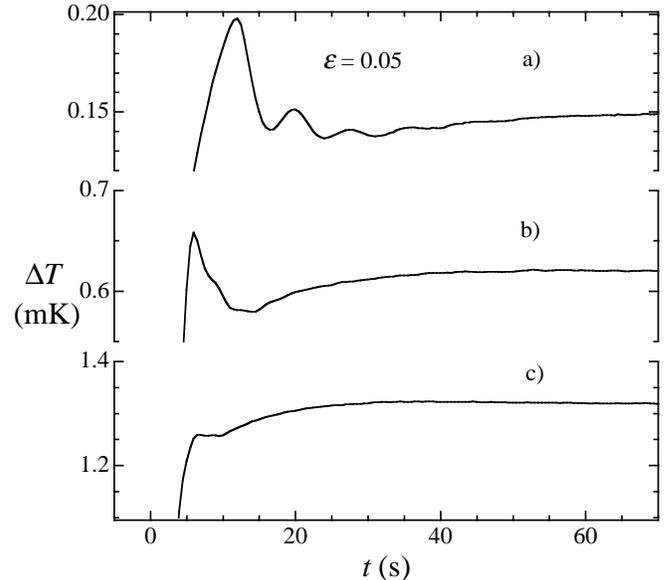,width=3.375in}}}
\caption{ Evolution of the profiles $\Delta T(t)$ at $\epsilon$ = 0.05 with
increasing values of $q$ (in $10^{-7}$ W/cm$^2$) : a) 9.65, b) 61.3, c) 154.}
\end{figure}

For the determination of $\tau_{tail}$, which ranged from $\sim$ 5 to 
100 s., the
recorded curve was fitted to an exponential, and a check of the 
dependence  on the
fit range showed an internal consistency in
$\tau_{tail}$ within about $\pm$ 10
$\%$. The relaxation profile distortion by the instrumental time 
constant has been
considered \cite{Schaeffer}. Besides an effective overall lag, there is no
significant change produced in the value of  $\tau_{tail}$. The 
oscillation period
$t_{osc}$ was obtained from the time count between succeeding minima, with an
average error of approximately $\pm$ 10$\%$. In this determination, 
the position of
the first maximum in $\Delta T(t)$ was not included. This was because the time
$t_m$ between this maximum  and the first minimum, as defined in
ref.\cite{Chiwata:O:2001}, is larger than half  of the oscillation period
$t_{osc}$. This asymmetry  is   most evident for
$\epsilon$ = 0.2, (See Fig.3a) and gradually decreases with
$\epsilon$.

At high enough values of $q$, an additional very weak and broad 
maximum is observed
before $\Delta T(\infty)$ is reached from above. The amplitude of 
this component is
only at most 0.3 $\%$ of $\Delta T(\infty)$ and the corresponding 
relaxation time
to steady state is estimated to be about O(5$\tau_{tail}$). In our analysis, we
have neglected this additional component.

\subsection {Region for $\epsilon \leq 9\times10^{-3}$}

As already described in ref. \cite{Kogan:M:2001}, the profiles of 
$\Delta T(t)$  in
this region are quite different from those at higher $\epsilon$. In 
zone E of Fig.
2, no overshoot is observed. After the start of the heat flow, the increase of
$\Delta T(t)$ is found to be consistent with diffusive heat
transfer\cite{Behringer:O:M:90} until the onset of instability leads 
to saturation
and to the plateau $\Delta T(\infty)$, as shown by the three lowest traces in
Fig.12a) of ref.\cite{Kogan:M:2001}. Above the passage from zone E to 
B, indicated
by a hatched band in Fig.2, there is a gradual appearance of an  overshoot with
increasing $q$, but without oscillations.

\section{Scaled times versus Rayleigh number}

Steady state measurements in both compressible and essentially incompressible
liquids
\cite{Ashkenazi:S:99,Niemela:2000,Chavanne:2001} show  approximate 
scaling of the
data on a $Nu^{corr}(Ra^{corr})$ plot leading to a universal curve. 
This is also
the case for supercritical $^3$He at various temperatures, but reported over a
larger range of
$\beta_T$ and $Pr$ than previously. The degree of scaling of
\underline{steady-state} data can be seen in Fig.7a and 8 of
ref.\cite{Kogan:M:2001} (See however note\cite{anomalous}). Is there 
similarly for
the
\underline{transient} data a  corresponding  representation of scaled 
times that
leads to a universal curve? This section  deals with this question.

Since data extend from the  regime of early turbulence ($Ra \sim 
5\times 10^6$) to
close to the convection onset, we express them in terms of the 
``distance" from this
transition, [$Ra^{corr}-Ra_c$]. It is useful here to present the 
relation with the
measured quantities $\Delta T$ and
$\Delta T_{ons}$. From a combination of  Eqs. 3 and 5, one obtains
\begin{equation}
     [Ra^{corr} - Ra_c] = B(\epsilon)[\Delta T - \Delta T_{ons}]
\end{equation}
where $B(\epsilon) \equiv \rho g \alpha_P L^3/(\eta D_T)$, with 
numerical values
showing a strong  dependence on $\epsilon$ along the critical isochore, as
presented in Table 1.

\begin{table}
\begin{tabular}{r|c|c|c|c|c|c} $\epsilon$  & $\beta_T$
(cm$^2$/dyne)  & D (cm$^2$/s)  & $\gamma$  & $Pr$ & $B[\epsilon]$
(K$^{-1}$) &  $t_D$(s)\\
\hline \hline 5.0e-4 &1.52e-3  & 7.3e-7 &3.0e3 &  590 & 7.0e12  &
  3.8e3 \\  1.2e-3 & 5.7e-4 & 1.4e-6 & 1.2e3 & 290 & 1.4e12
&  2.0e3\\ 2.0e-3 & 2.91e-4 & 2.3e-6 & 6.6e2 & 180 & 4.3e11  &
  1.2e3\\ 5.0e-3 & 9.61e-5 & 5.5e-6 & 2.5e2& 75 & 6.1e10  &
  5.1e2 \\ 1.0e-2 & 4.30e-5 & 1.1e-5 & 1.2e2 & 38 & 1.4e10 &
  2.6e2\\ 2.0e-2 & 1.88e-5 & 2.2e-5 & 58 & 19 & 3.1e09 &
1.3e2 \\  5.0e-2 & 6.32e-6 & 5.4e-5 & 23 & 7.4 & 4.2e08  &  52 \\
1.0e-1 & 2.77e-6 & 1.1e-4 & 12& 3.8 & 9.8e07   & 27 \\
2.0e-1 & 1.21e-6 & 1.9e-4 & 6.5 & 2.1 & 2.4e07  &  15\\
\hline
\end{tabular}
\caption{Several properties (numbers rounded off) for $^3$He along its
critical isochore, relevant to the convection onset transient.
$B[\epsilon]$, defined in Eq.6
   and $t_D = L^2/4D$ are functions of the layer thickness
$L$ and have been calculated for the present cell, $L$ = 0.106 cm.}
\end{table}

We have found that  both $t_{osc}$ and
$\tau_{tail}$, when scaled by the diffusive time $t_D= L^2/4D_T$
\cite{diffusivetime}, and plotted versus  [$Ra^{corr}-Ra_c$], can be 
collapsed on
two separate curves. The range of scaling extends over about 3.5 decades in $[Ra^{corr}-Ra_c]$.  In Fig.5 we show this scaling plot where the 
various symbols
indicate the temperatures $\epsilon$ at which the transients were
recorded\cite{table}. There are some systematic deviations and 
scatter of less than
$\pm$ 15\% from the average curves.

   This is a  remarkable result because with decreasing $\epsilon$ from 0.2 to
$9\times 10^{-3}$,  the isothermal compressibility
$\beta_T$  and
$t_D$ increase by factors of $\approx$40  and 20 respectively , and the Prandtl
number
$Pr$ increases from 2 to 42. Over approximately 2.5 decades of 
[$Ra^{corr} - Ra_c$],
the  curves for
$t_{osc}/t_D$ and $\tau_{tail}/t_D$  can be represented by power laws with
exponents of
$\simeq$ -0.52 and -0.60, as obtained from a ``best" straight line through the
points. Because of the non uniform data distribution over the range of
Rayleigh numbers, a least-square fit seems inappropriate. Both scaled times
diverge as [$Ra^{corr}-Ra_c]
\rightarrow$ 0 and there appears to be no clear dependence on $Pr$, 
as shown by the
collapsing of the data for various values of
$\epsilon$. We can therefore expect that the representations in Fig.5 
should be the
same for other fluids kept at constant volume, and hence they form in 
a crude first
approximation a universal set of curves. We note that for convection transients
observed at constant pressure in liquid $^4$He
\cite{Behringer:85}, $\Delta T(t)$ relaxes from its ``overshoot" 
maximum towards
$\Delta T(\infty)$ with a characteristic time that diverges as [$Ra-Ra_c]
\rightarrow$ 0, similarly showing the ``slowing down" as the 
transition to fluid
stability is approached.

\begin{figure}[tb]
\center{\parbox{3.375in}{\psfig{file=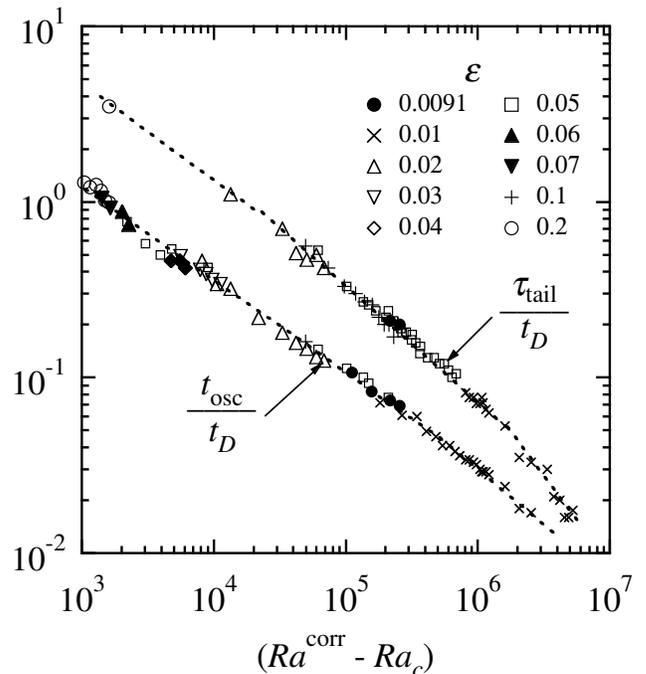,width=3.375in}}}
\caption{Scaled plots of $t_{osc}/t_D$ and $\tau_{tail}/t_D$ versus
[$Ra^{corr}-Ra_c$] for reduced temperatures 9$\times 10^{-3} \leq \epsilon
\leq $0.2. The dashed lines are ``guides to the eye".}
\end{figure}

As mentioned before, no damped oscillations were detected for 
$\epsilon < 9\times
10^{-3}$. To check whether this absence could be caused by the 
instrumental time
constant if $t_{osc}$ becomes too short, a calculation of the 
expected $t_{osc}$ was
made based on the scaling plot of Fig.5. This calculation was done 
for $\epsilon =
5\times 10^{-3}$ and $3 \times 10^{-3}$ and for several values of 
$\Delta T$ where
experiments had been  carried out. For
   these $\Delta T(t)$ profiles, the predicted $t_{osc}$ values were 
of order 15 s.,
large enough to be clearly observable. Hence we conclude that the 
character of the
transient has changed when $\epsilon$ is below 9$\times 10^{-3}$. 
Furthermore  the
$\tau_{tail}/t_D$ obtained from analysis of the $\Delta T(t)$ recordings at
$\epsilon$ = 5$\times 10^{-4}$ were found to be inconsistent with an 
extrapolation
from the scaled plot in Fig.5 to the higher
$Ra$ numbers ($\sim 4\times 10^8$) corresponding to these data. The 
steady-state
data at this temperature also showed an anomalous behavior in the
$Nu^{corr}(Ra^{corr})$ plot\cite{Kogan:M:2001}.

It has been shown in several papers
\cite{Chiwata:O:2001,Amiroudine:Z:2002,Furukawa:O:2002} from 
simulations that the
compressible fluids should give oscillatory  $\Delta T(t)$ 
arbitrarily close to the
convection onset.  This behavior is associated with the formation of 
the boundary
layers responsible for the ``Piston Effect" and thus should occur in 
compressible
fluids kept at constant volume, but not at constant pressure. The 
boundary layers
are predicted to undergo an instability which leads to a sequence of 
``plumes" of
hot and cold fluid  moving across the cell periodically, and 
producing oscillations
in
$\Delta T(t)$. Eventually the oscillations disappear as the ``plumes" 
give way to
the time-independent convection rolls.

\section{Comparison of experiment with  predictions}

In this section we are concerned with the time $t_m$ between the 
first peak and the
first minimum of the $\Delta T(t)$ profile (See  Figs 1 and 3). Specific
predictions for
$t_m$ have been  made analytically
\cite{Chiwata:O:2001} as well as obtained  from numerical simulations
\cite{Chiwata:O:2001,Furukawa:O:2002}. We extend the definition of 
$t_m$ to include
the region of truncated oscillation, as shown in Fig 1b and 3b, where 
the minimum
has been replaced by a kink followed by a plateau. This truncated overshoot is
found in Zone A of Fig.2. When plotted versus [$\Delta T - \Delta T_{ons}$] or
[$Ra^{corr} - Ra_c$], the line of  the $t_m$ data points for a given 
$\epsilon$ is
continued  smoothly from zone B into zone A.  Here we note that in neither the
simulations of ref\cite{Chiwata:O:2001} nor those of 
ref\cite{Amiroudine:Z:2002}
is  an evidence of the ``truncated oscillations" as was observed experimentally
(zone A). In the simulations, damped oscillations were obtained with 
a diverging
period  as
$[\Delta T - \Delta T_{ons}]\rightarrow$ 0.

\subsection{comparison with numerical simulations}

In Fig.6 we show the time $t_m$ as a function of [$\Delta T-\Delta 
T_{ons}$], and
the comparison  between the experiments and the 
simulations\cite{Chiwata:O:2001} at
$\epsilon$  = 0.05 for a range of heat currents $q$. As per Table 1,
$[\Delta T-\Delta T_{ons}] = 2\times 10^{-5}$ corresponds to 
$[Ra^{corr}-Ra_c] =
8.5\times10^3$. It can be seen that both sets of points give  almost 
parallel lines, with

\begin{figure}[htb]
\center{\parbox{3.375in}{\psfig{file=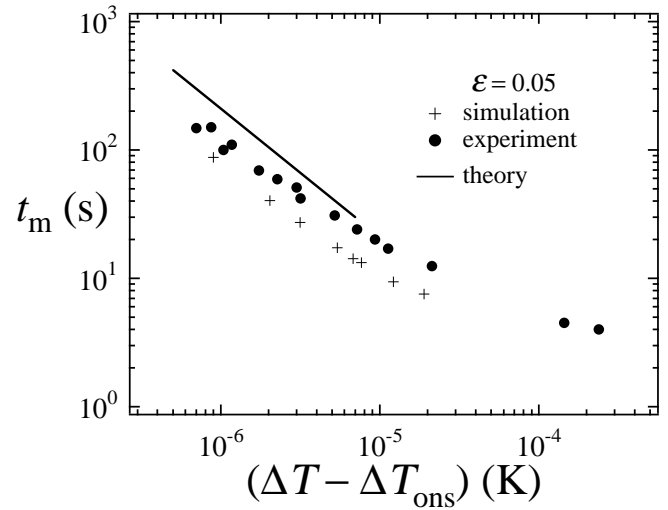,width=3.375in}}}
\caption{Logarithmic presentation of $t_m$ versus $[\Delta T$-$\Delta T_{ons}]$ at $\epsilon = 0.05$, from experiments and from simulations by Chiwata and Onuki [15]. The solid line represents the predicted behavior in the limit of small $[\Delta T - \Delta T_{ons}]$, Eq. 8.}
\end{figure}

\noindent  the amplitude difference being of the order of 30\%. In the 
limit of small heat
currents (or small values of [$\Delta T-\Delta T_{ons}$]), the exponent of both
lines of points tends to (-1). At high values of [$\Delta T-\Delta 
T_{ons}$] or of
[$Ra^{corr}-Ra_c$] the exponent is found to be
$\approx$ 0.5, and this result is similar to that found in Fig.5 for the scaled
$t_{osc}/t_D$ plot.

A simple prediction for the value of the limiting slope can be obtained in the
Stokes approximation, when [$Ra^{corr}/Ra_c -1]\leq Pr(\epsilon) $
\cite{Chiwata:O:2001}.  In this case, $t_m$ is estimated by setting 
$t_m \approx
L/v$, the time it will take for the cooler fluid near the top of the 
cell to flow
to the lower boundary. Here $v$ is the velocity estimated from arguments given
     under Eq.7 in ref.\cite{Chiwata:O:2001}, and $t_m$ is then given by

\begin{equation}
     (t_m/t_D)[Ra^{corr}/Ra_c -1]  = O(1)
\end{equation}
and, after combining with Eq.6, by
\begin{equation}
     t_m = O(1)t_DB(\epsilon)^{-1}[\Delta T - \Delta T_{ons}]^{-1}
\end{equation}
  with O(1) a numerical factor of the order of 1.
   (In ref.\cite{Chiwata:O:2001} the `` = O(1) " was written as `` 
$\approx$ ". See
also ref\cite{Furukawa:O:2002}, paragraph under Eq.3.8). This 
estimation, where we
have set O(1) = 1 is shown by a solid line in Fig. 6 for $\epsilon$ = 0.05. The
agreement between the estimations, the experiments and the 
simulations is within
the expected uncertainty of the estimation.

The simulations have given impressive agreement with the experimental 
results under
the steady-state conditions, for instance in the plot of
$\Delta T$ versus
$q$, (Fig.4 in ref\cite{Chiwata:O:2001}), in the convection current versus
$Ra^{corr}$,  Fig. 4 of ref\cite{Furukawa:O:2002} and also in the plots of
$\Delta T(t)$ at $\epsilon$ = 0.01 by AZ\cite{Amiroudine:Z:2002}. While on
the whole the agreement between simulations and experiments for the oscillatory
transient is good, we note some minor  inconsistencies. From the 
simulation plots,
Figs.1 and 2 in ref.\cite{Chiwata:O:2001}, Fig. 1a in 
ref.\cite{Furukawa:O:2002},
both at 0.05, and similarly from plots by AZ at $\epsilon$ = 0.01,  the
ratio
$t_{osc}/2t_m$ is always $>$ 1. By contrast, in the experiments the 
ratio is $<$ 1,
as can be seen from Figs. 1a, 3a and 4a in this paper. Also in the 
simulations of
ref~\cite{Furukawa:O:2002}, Fig.1a shows  no evidence of the observed slow
relaxation to the steady-state value of $\Delta T$. Here the simulations are
compared with the same experimental data as in  Fig.4a of the present paper.
Hence, so far simulations do not provide a prediction for
$\tau_{tail}$. Also,  as mentioned before, they do not show truncated 
oscillations
(zone A in Fig.2).

\subsection{Comparison with theory}

We have first examined  for the range $ 0.009 \leq \epsilon \leq $0.2 a plot of
$t_m/t_D$ versus [$Ra^{corr} - Ra_c$] similar to Fig. 5.  The
$t_m$ data include results from both zones A and B, and extend to a 
lower range of
[$Ra^{corr} - Ra_c$] than in Fig.5. However, in contrast to the
$t_{osc}/t_D$ plot, we  found that the data did not fall on a single curve. The
spread is particularly large at low values of [$Ra^{corr}-Ra_c$]. Hence such a
scaled representation was not found informative. Instead,  for the 
comparison of
the prediction with experiments, it was
   found  more convenient to formulate
$t_m$ in terms of temperature differences [$\Delta T - \Delta T_{ons}$] than of
  $[Ra^{corr} - Ra_c]$, as it gives a more compact data representation.

\begin{figure}[htb]
\center{\parbox{3.375in}{\psfig{file=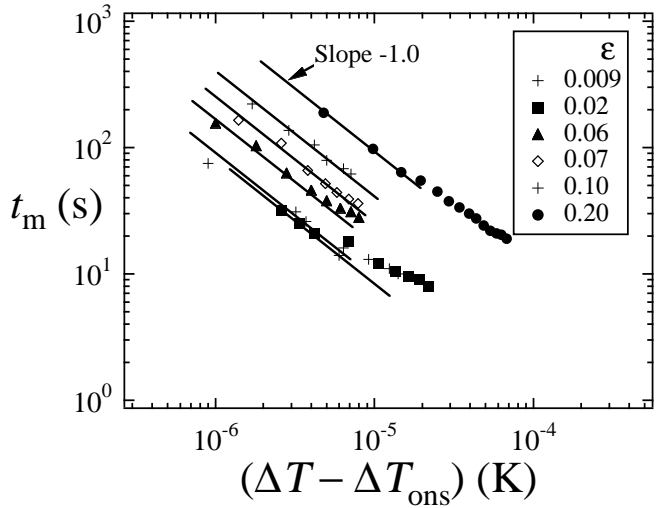,width=3.375in}}}
\caption{Experimental data of $t_m$ versus  [$\Delta T - \Delta T_{ons}$] for
selected reduced temperatures in the range
$9\times 10^{-3} \leq \epsilon \leq 0.2$. To avoid overcrowding, half 
of the curves
at constant $\epsilon$ have been suppressed. The solid lines, all with slope of
(-1) are the result of a fit to the data in the limiting region of 
small values of
[$\Delta T -
\Delta T_{ons}$]. The slope (-1) of the solid line
   represents the predicted behavior.}
\label{fig:}
\end{figure}

In Fig.7 we show   experimental data sets  at selected values of
$\epsilon$, to avoid overcrowding. (The set for $\epsilon$ = 0.05, 
shown in Fig.6,
has not been included here) and all were fitted  with straight lines 
of slope of
(-1) in the limit of small values of [$\Delta T - \Delta T_{ons}$]. 
These fits gave
the corresponding amplitudes, which were converted into amplitudes of 
[$Ra^{corr} -
Ra_c$] by means of Eq. 6. The result is presented in Fig.8  in the 
scaled form of
Eq. 7 as  a function of $\epsilon$. Since the expected ratio is of order unity
\cite{Chiwata:O:2001}, there is agreement with predictions at the 
higher values of
$\epsilon$, but a systematic deviation appears as
$\epsilon$ decreases below 0.02, perhaps anticipating the crossover 
to the regime
$\epsilon \leq 9\times10^{-3}$, where damped oscillations are no 
longer observed.

\begin{figure}[htb]
\center{\parbox{3.375in}{\psfig{file=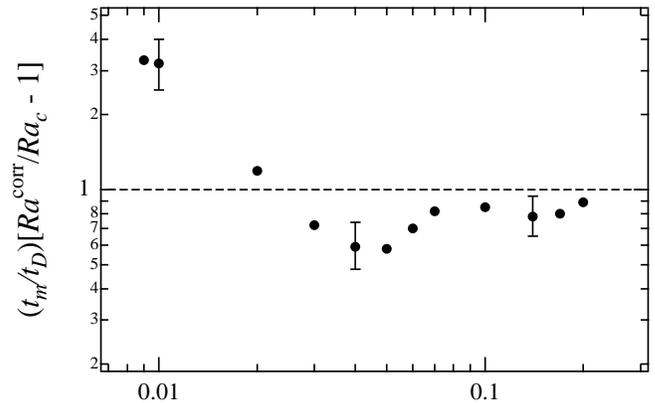,width=3.375in}}}
\caption{The ratio $(t_m/t_D) [Ra^{corr}/Ra_c - 1]$ from Eq.7 versus 
the reduced
temperature
$\epsilon$. The symbols (solid circles) represent the results obtained from the
plot in Fig. 6 and 7 including all the values of $\epsilon$ where 
convection data
were taken. According to the 
predictions\protect\cite{Chiwata:O:2001}, the ratio
should be of the  order of 1 and independent of $\epsilon$.}
\end{figure}

\section{Discussion and Summary}

We have analyzed the time-dependent temperature difference $\Delta 
T(t)$ across the
fluid layer along the critical isochore in a Rayleigh-B\'{e}nard cell 
with a height
of
$L$ = 0.1 cm and a large aspect ratio. The experiments studied the 
behavior of the
fluid in its convective state from the start of a constant heat flow and until
steady state was observed. The measurements were done with  supercritical
$^3$He along the critical isochore where the physical properties change rapidly
with the distance from $T_c$. A characteristic time for this system is
$t_D = L^2/4D$, the diffusive time, where $D_T$ is the thermal diffusivity. The
experiments have shown profiles of $\Delta T(t)$ gradually evolving 
with reduced
temperature $\epsilon$ and heat flow $q$. The principal results can 
be summarised
as follows.

1) The transient profiles appear to fall into two temperature regimes:  above
and  below $\epsilon = 9 \times 10^{-3}$. In the regime above,
damped oscillations are observed, but not so in the regime below.
The oscillations have subsequently  been reproduced in simulations by 
Onuki and his
group and by Amiroudine and Zappoli at $\epsilon$ = 0.05 and 0.01.

2) For $\epsilon \geq 9 \times 10^{-3}$, and for increasing heat 
flows, there is
first a region of truncated oscillations. For larger  heats, damped 
oscillations
with a period
$t_{osc}$ are observed, and at still larger heat also a tail of the 
transient which
tends to the steady-state value $\Delta T$ with a relaxation time 
$\tau_{tail}$.
It is found that the scaled values of $t_{osc}/t_D$ and 
$\tau_{tail}/t_D$ fall on
two separate curves of points when plotted versus the Rayleigh number. This is
remarkable, considering that the physical properties change very 
rapidly  along the
critical isochore.

3) For low Rayleigh numbers, predictions were made by Onuki and 
coworkers for the
time $t_m$ between the first maximum and the following minimun in the 
oscillatory
transient $\Delta T(t)$. We find their results to be in  approximate 
agreement with
our data.

4) While there is an impressive semiquantitative agreement between the observed
$\Delta T(t)$ profiles, and those simulated in 2D under identical 
conditions of $L$
and fluid properties, there are some differences in detail. This 
might be due to
the nature of the simulations where the geometry is different  from that of the
experimental cell. First, the simulations did not reveal a regime of truncated
oscillations, as did the experiments. Second, there are subtle, but 
nevertheless
visible differences in the oscillation patterns as shown by the  ratios
$t_{osc}/2t_m$ which are $>$ 1 in the simulations but generally $<$ 1 in the
experiments.

5) For $\epsilon < 9\times 10^{-3}$ no truncated or damped oscillations were
detected while an extrapolation of the observations above this regime indicates
that oscillations should have been observable. Hence the character of 
the profile
$\Delta T(t)$ appears to be different in this temperature regime, and 
needs to be
understood.

\section{Acknowledgments} One of us (HM) greatly acknowledges stimulating
discussions and correspondence with A. Onuki, S. Amiroudine and P. 
Carl\`{e}s. We
thank A. Onuki for communicating the simulation data used in Fig.6. 
The  efficient
and willing  help of F. Zhong and M. Gehm in  formatting the figures is greatly
appreciated. We also thank K. Sreenivasan and J. Niemela for their very
appropriate suggested improvements in the text. This work has been 
supported by NASA
grant No NAG3-1838.


\begin{references}

\bibitem{Tritton:88} D.J. Tritton, Physical Fluid Dynamics (Oxford 
Science, Oxford,
1988) Chapter 14.

\bibitem{Kadanoff:2001}L.P. Kadanoff, Physics Today\ {\bf 54}, \# 8, 34 (2001).

\bibitem{Ashkenazi:S:99} Sh. Ashkenazi and V. Steinberg, Phys.\ Rev.\ 
Lett {\bf83},
3641 (1999).

\bibitem{Niemela:2000} J.J. Niemela, L. Skrbek, K.R. Sreenivasan and 
R.J. Donnelly,
Nature {\bf404}, 837 (2000).

\bibitem{Chavanne:2001} X. Chavanne, F. Chilla, B. Chabaud, B. Castaing and B.
Hebral, Phys.\ Fluids {\bf13}, 1300 (2001).


\bibitem{Landau:L:}L.D. Landau and E.M. Lifshitz, Course of 
Theoretical Physics:
Vol.6 Fluid Mechanics (Pergamon, Oxford, 1959)

\bibitem{Kogan:M:2001} A.B. Kogan and H. Meyer,  Phys.\ Rev.\ E {\bf 
63 }, 056310
(2001).

\bibitem{Onuki:F:90} A. Onuki and R.A. Ferrell, Physica A {\bf 64}, 245 (1990).

\bibitem{Beysens} P. Guenoun, B. Khalil, D. Beysens, Y. Garrabos, F. 
Kammoun, B. Le
Neindre and B. Zappoli,  Phys.\ Rev.\ E {\bf 47},1531 (1993); H. Boukari, J.N.
Shaumeyer, M.E. Briggs amd R.W. Gammon, Phys.\ Rev.\ A {\bf 41}, 2260 
(1990); H.
Klein, G. Schmitz and D. Woermann, Phys.\ Rev.\ A {\bf43}, 4562 
(1991); F. Zhong
and H. Meyer, Phys.\ Rev.\ E {\bf 51}, 3223 (1995); J. Straub, L. Eicher and A.
Haupt, Intern.\ J.\ Thermophysics {\bf16}, 1033 (1995); A. Kogan and 
H. Meyer, J.\
Low Temp.\ Phys.\ {\bf112}, 419 (1998).

\bibitem{adiabatic} For a simple meaning of $\nabla T|_{ad}$ as $T_c$ 
is approached
and
$C_P/C_V\gg$1, see ref.\cite{Kogan:M:2001} below Eq. 2. Under these conditions,
$\nabla T|_{ad}$ tends to the constant value of 34
$\mu$K/cm for $^3$He.

\bibitem{Gitterman:S:71} M. Gitterman and V. Steinberg, J. Appl. 
Math. Mech. USSR
{\bf34}, 305 (1971). M. Gitterman, Rev. Mod. Phys. {\bf 50}, 85 (1978).

\bibitem{Carles:U:99} P. Carl\`{e}s and B. Ugurtas, Physica D, {\bf 
126}, 69 (1999).

\bibitem{Amiroudine:2000} S. Amiroudine, A.B. Kogan, H. Meyer and B. 
Zappoli, Proc.
Internat.  Congress  Theor\& Appl. Mechanics, (ICTAM) Chicago Aug. 
27, 2000, edited
by H. Aref and J. W. Phillips, Kluwer Academic Publishers (2001).

\bibitem{Amiroudine:Z:2002} S. Amiroudine and B. Zappoli (private 
communication,
and to be published).


\bibitem{Chiwata:O:2001}Y. Chiwata and A. Onuki, Phys.\ Rev.\ Lett.\ {\bf 87},
144301 (2001).

\bibitem{Furukawa:O:2002} A. Furukawa and A. Onuki Phys.\ Rev.\ E {\bf 66},
  016302 (2002).

\bibitem{Carles:2000} P. Carl\`{e}s, Physica D {\bf147}, 36 (2000).

\bibitem{Carles:2002} L. El Khoury and P. Carl\`{e}s, (to be published).


\bibitem{Behringer:O:M:90} R.P. Behringer, A. Onuki and H. Meyer, J.\ 
Low Temp.\
Phys.\ {\bf81}, 71 (1990).


\bibitem{Behringer:A:77} R.P. Behringer and G. Ahlers, Phys. Lett. 
{\bf A 62}, 329
(1977).

\bibitem{Behringer:85} R.P. Behringer, Rev. Mod. Phys. {\bf57}, 657 (1985).


\bibitem{Schaeffer} We are indebted to D. Schaeffer of the Duke U. Mathematics
Department for  producing an equation expressing the observed $\Delta 
T(t)$ signal
in terms of an input signal modified by an instrumental time constant 
shorter than
$\tau_{tail}$.

\bibitem{anomalous} The anomalous behavior observed at the highest values of
$Ra^{corr}$ is not understood and will not considered further here.

\bibitem{diffusivetime} This diffusion time is as defined in
ref.\cite{Chiwata:O:2001}, and differs from the usual diffusive time 
$L^2/D_T$ in
ordinary fluids by a factor of 4. This is because $t_D$ is chosen for 
the situation
of a supercritical fluid where
$C_P>>C_V$, and for relaxation at constant density. Under such conditions, the
relaxation time is 4 times shorter than when $C_P \approx C_V $ ( See
ref.\cite{Behringer:O:M:90})

\bibitem{table} A tabulation of scaled values of $t_{osc}$ and $\tau_{tail}$ is
available from one of us (HM) for the temperature range 9$\times10^{-3} \leq
\epsilon \leq$0.2. Here the relevant values of $t_D$ are given for each
$\epsilon$.


\end{references}
\end{document}